\documentclass[fleqn,10pt]{wlscirep}
\usepackage[utf8]{inputenc}
\usepackage[T1]{fontenc}
\title{Misaligned orientations of 4f optical neural network for image classification accuracy on various datasets}

\author[1]{Yanbing Liu}
\author[2,*]{Wei Li}
\author[1]{Kun Cheng}
\author[2]{Xun Liu}
\author[1,+]{Wei Yang}
\affil[1]{School of Electronic Engineering, Beijing University of Posts and Telecommunications,Beijing,100876,China}
\affil[2]{Beijing Institute of Space Mechanics and Electricity,China Academy of Space Technology, Beijing,100094,China}
\affil[*]{wei$\_$li$\_$bj@163.com}
\affil[+]{yangwei@bupt.edu.cn}

\begin{abstract}
In recent years, the optical 4f system has drawn much attention in building high-speed and ultra-low-power optical neural networks (ONNs). Most optical systems suffer from the misalignment of the optical devices during installment. The performance of ONN based on the optical 4f system (4f-ONN) is considered sensitive to the misalignment in the optical path introduced. In order to comprehensively investigate the influence caused by the misalignment, we proposed a method for estimating the performance of a 4f-ONN in response to various misalignment in the context of the image classification task.The misalignment in numerical simulation is estimated by manipulating the optical intensity distributions in the fourth focus plane in the 4f system. Followed by a series of physical experiments to validate the simulation results. Using our method to test the impact of misalignment of 4f system on the classification accuracy of two popular image classification datasets, MNIST and Quickdraw16. On both datasets, we found that the performances of 4f-ONN generally degraded dramatically as the positioning error increased. Different positioning error tolerance in the misalignment orientations was observed over the two datasets. Classification performance could be preserved by positioning errors up to 200 microns in a specific direction.
\end{abstract}
\begin{document}

\flushbottom
\maketitle

\thispagestyle{empty}

\section*{Introduction}
The emergence of deep convolutional neural networks started the third boom in artificial intelligence research\cite{a1,a2,a3,a4,a5,a6,a7,a8,a9}. In recent years, with the rapid development of deep convolutional neural networks in image processing, image recognition and image generation, which have been widely applied in the applications of biomedicine, aerospace and intelligent driving\cite{a10,a11,a12 ,a13,a14,a15,a16,a17,a18,a19,a20,a21} and achieved many promising results. Leveraging the continuous research attention on the convolutional neural networks, the scale of convolutional neural networks is also increasing exponentially \cite{a1,a22,a23,a24,a25,a26}. However, the ultra-high power consumption of the von Neumann architecture\cite{a27} is now the bottleneck of the deep neural networks, which severely limits the application of deep neural networks in edge fields such as unmanned driving and intelligent industry. Consequently, the optical neural network (ONN) computing architecture attracts more attention because it uses light as the computing carrier, significantly reducing power consumption. 
ONN based research \cite{a28,a29,a30,a31,a32,a33,a34,a35,a36,a37,a38,a39,a40,a41,a42,a43,a44,a45,a46,a47,a48,a49}, including pre-optical neural network systems, recurrent neural network optical computing systems and impulse neural optical computing systems. The pre-optical neural network architecture is considered as the most feasible solution for deployment to edge applications \cite{a40,a50,a51,a52,a53,a54,a55,a56,a57,a58,a59,a60,a61}, the linear approximation of ONN is currently the relatively mature implementation method, such as diffractive optical neural network architecture\cite{a50,a51,a52,a53,a54,a55,a56,a57} and optical 4f-based systems \cite{a58,a59,a60, a61} etc.

Due to the simple and straight organization, the optical 4f system is the primary way to construct the linear convolution optical computing layer in the pre-optical neural network. While the optical 4f system modulates the light field of the input image spectrum at its second focus (2f), so it cannot directly convolve the input image with multiple convolution kernels superimposed in depth like a computer neural network (CNN) does. And the optical 4f system is quite sensitive to misalignment errors, a small misalignment error of  optical system will affect the final output of the system.Therefore, in order to evaluate the effect of misalignment error on the final output of the optical 4f system, we perform numerical calculations and physical experiments to analyze the effect of the misalignment at 2f on the output image of 4f system.

The structure of this paper is organized as follows. In section 2.1, we design a set of data sets constructed by four simple binary plots. The LFMM parameters are directly obtained using machine learning end-to-end training. We verified the feasibility and accuracy of the LFMM using both theoretical simulation and physical experiments. It is found that the mismatch between the light field modulation device at 2f and the spectral light field distribution of the input image will seriously distort the light field distribution of the output image of the 4f-ONN.
In order to further explore the influence of the mismatch on the classification results, in Section 2.2 of the paper, we select the MNIST and Quickdraw16 dataset recognized in the field of deep learning for research.
Through a series of simulations, we found that the classification accuracy of different datasets has an evident orientation dependence on the mismatch direction of the light field modulation device. The MNIST dataset is sensitive to the mismatch deviation in the horizontal axis, while the Quickdraw16 dataset is entirely different, which is more sensitive to mismatch extent in the vertical axis.
We qualitatively analyzed the possible explanations for this observation regarding the image distribution characteristics of the two datasets respectively;in the method section of the paper, we give the theoretical basis and training process of the proposed LFMM base 4f-ONN.

\section*{Results}
\subsection*{Theoretical simulation and experimental verification}

\begin{figure}[h]
	\centering
	\includegraphics[width=\linewidth]{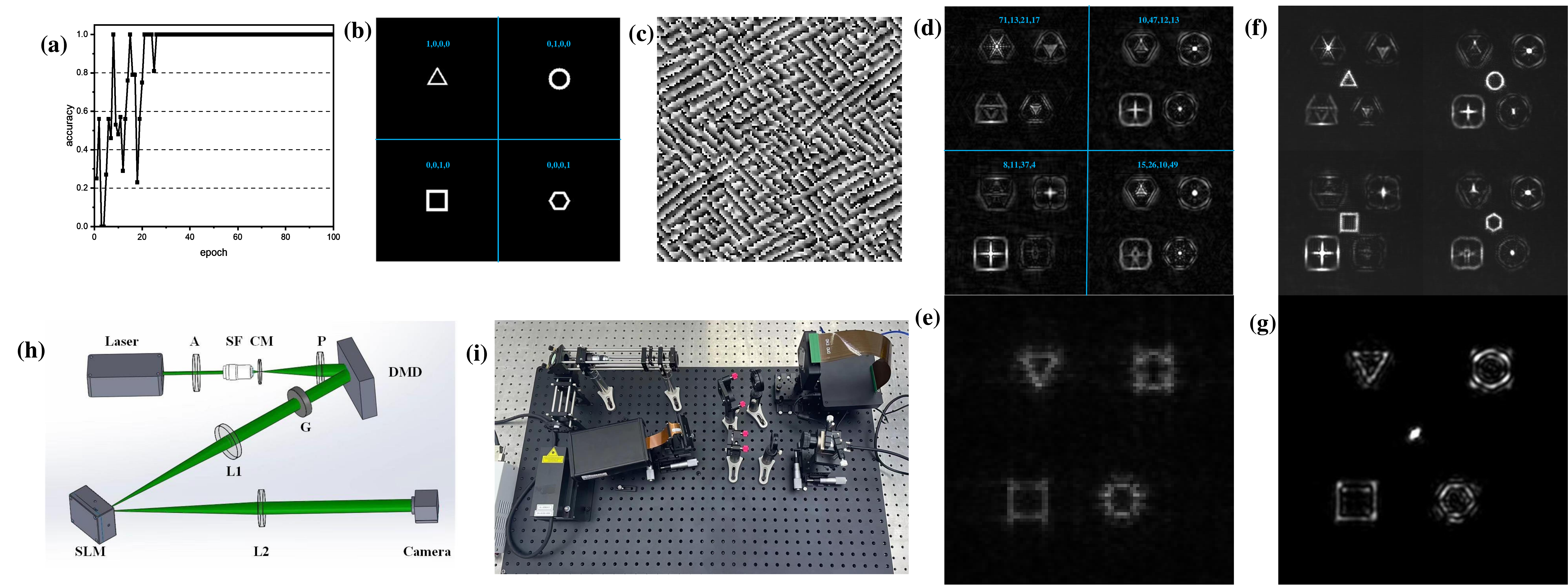}
	\caption{(a)Training accuracy on the toy image.(b)Input image one hot label. (c)Phase map at 2f.(d)Output images of four simple plots (from top left to bottom right : triangle, circle, square and hexagon.(e)The corresponding PSF of trained LFMM.(f)PSF of 4f-ONN.(g)4f-ONN outputs of the 4 shapes respectively.(h)Experiment workflow.(i)Physical installation of the experiment}
	\label{1}
\end{figure}

In order to evaluate the effect of misalignment error on the final output of the optical 4f system, we designed a set of simulations and physical experiments. First, to accurately align the optical path and calibrate the parameters of each optical device, we made a toy image consisting of four synthetic plots, including a circle, a triangle,  a square and a hexagon, as shown in Fig. \ref{1}(b).
As described in the method,  the LFMM that can classify these four simple graphs is pre-trained. Overfitting occurs as we expected due to the 1 image only training set, resulting 100\% classification accuracy after around 20 epochs as shown in Fig. \ref{1}(a).

The LFMM obtained by training is calculated by the exponential function of the phase map at 2f, as shown in Fig. \ref{1}(c), and the corresponding PSF of the 4f system is shown in Fig. \ref{1}(e). 
For simplicity, we feed the image of four simple plots ensembled as 2 × 2 array into the trained  convolutional neural network based on 4f(4f-CNN). Fig. \ref{1}(d) shows the output of the 4f system,  listed 2 × 2  in the same order in Fig. \ref{1}(b). 
It can be seen that the light field intensity distribution of the output image at 4f of different plots has noticeable geometric structure differences and the intensity peaks at the corresponding location of the 2 × 2 alignment. For example, we give one hot label in blue in Fig. \ref{1}(b).
When the input image is a triangle (1,0,0,0), the maximum pooling of Fig. \ref{1}(d) outputs the top left segment (71,13,21,17), which gives the correct class prediction.
Experiments based on synthetic images show that the output results at 4f have clear outlines and visible differences, which indicates significantly discriminative domain boundaries.
Next, we will build a physical optical path and transfer the 4f-CNN model to 4f-ONN to demonstrate the practical feasibility of our method and to 
check the consistency of 4f-ONN and 4f-CNN output images.

The flow chart and physical installation of the experiment are shown in Fig. \ref{1}: A stands for an adjustable attenuator, which is used to adjust the strength of the laser;
SF is a beam expander spatial filter, which expands the laser beam emitted by the laser so that it can completely cover the Digital Micromirror Devices(DMD) to ensure the integrity of the input image;
CM is a collimating lens, which is used to ensure that the laser direction is horizontal. P is a polarizer, which filters out longitudinal waves in the laser; G is a diaphragm, which is used to limit the size of the beam; L1 and L2 are two lenses with a focal length of 200  mm, performing Fourier transform and inverse transformation in the 4f system,respectively.
The spatial light field modulator (SLM) modulates the spectral light field of the input image spectrum at 2f. The SLM used is a pure phase spatial light modulator called LETO manufactured by Holoeye, Germany, with a resolution of 1920 x 1080, a pixel pitch of 6.4 µm, a small interpixel gap of 0.2 µm, provides a high fill factor of 93\%, and the phase modulation range is 0-2$ \pi $(0-255);
The input image is loaded by a DMD, which is DLPC900 chip by Texas Instruments. Its resolution is 1920 x 1080, the pixel size is 7.56 µm, the micromirror tilt angle is $ \pm12^{\circ} $, and the fill factor is 92$\%$. The camera at 4f receives the output image is a MER-130-30-UM by Daheng Imaging, the sensor is progressive scan CMOS, the horizontal/vertical resolution is 1280 x 1024, the pixel size is 5.2 µm x 5.2 µm, and the frame rate is 30 fps.

The emitted laser passes through light intensity adjustment, beam expanding, collimating and polarizing, and then incident on the DMD as a parallel coherent shear wave. 
Finally, the light beam modulated by the SLM is received by the camera through lens 2, which is the final system ouput parameterized by Eq. (\ref{104}).

Based on the specification of the experimental devices, we adjusted the input image and the pre-trained grayscale images to appropriate sizes before loading them on the DMD and SLM. 
The PSF of the 4f-ONN and the output image at 4f obtained from the experiment are shown in Fig. \ref{1}(g) and Fig. \ref{1}(f), respectively.

Overall,  the experimental and theoretical simulation results are consistent. The 4f-CNN and 4f-ONN can also produce consistent results for different input shapes, and the performance is stable, which verifies the feasibility and robustness of our proposed method and the consistency of 4f-ONN and 4f-CNN output images.

We are curious about the systemic response of the 4f-ONN system when the input image spectrum does not match the size and spatial location of the phase map loaded on the SLM. Taking the circle plot input as an example, we investigate the mismatch in size and spatial alignment of the input image spectrum and SLM, respectively.

\begin{figure}[h]
	\centering
	\includegraphics[width=\linewidth]{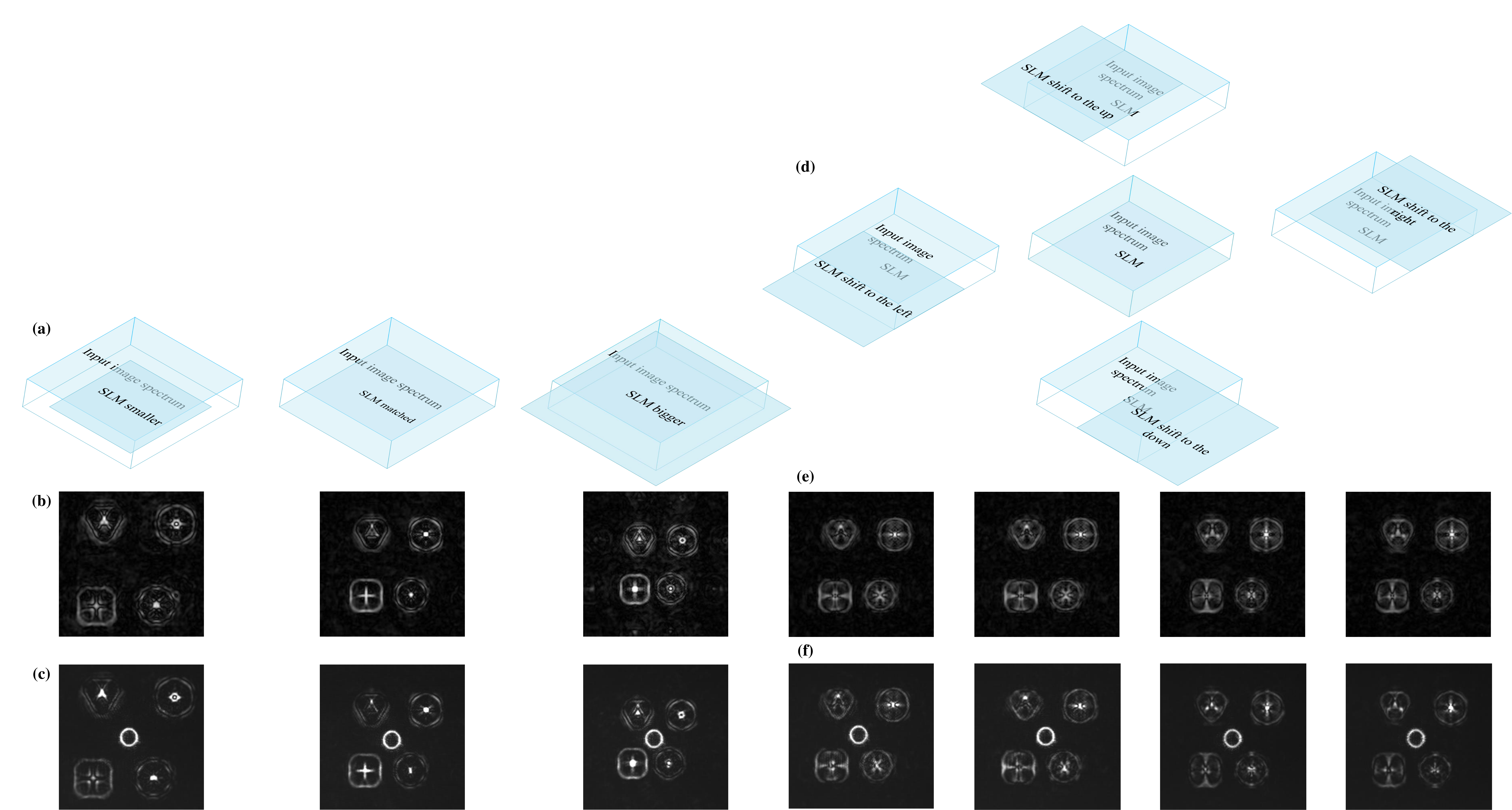}
	\caption{(a)Scale mismatch experiments.(b)Scale mismatch experiments on 4f-CNN.(c)Scale mismatch experiments on 4f-ONN.(d)Alignment mismatch experiments.(e)Alignment mismatch experiments on
4f-CNN.(f)Alignment mismatch experiments on 4f-ONN}
	\label{2}
\end{figure}

The schematic diagram of the experiment about scale mismatch is shown in Fig. \ref{2}(a), from left to right, the size of the phase image loaded on the SLM is less than, exactly equal to, and greater than the size of the SLM at 2f, simulation and experimental results are shown in Fig.\ref{2}(b) and Fig. \ref{2}(c), respectively.

The schematic diagram of the experiment of spatial alignment mismatch is shown in Fig. \ref{2}(d), the SLM is shifted to the left, right, up, and down by 18 pixels (700 microns) away from the perfect match, respectively. The simulation and experimental results are shown in Fig. \ref{2}(e) and Fig. \ref{2}(f), respectively.

White noise is observed in the background of the output when SLM does not match the size of the input image.  When the SLM is smaller than the input, the four figures of the output image tend to scatter outward; Conversely, the four figures of the output image tend to converge inward.
In terms of spatially alignment mismatch, the output of the 4f-ONN and 4f-CNN are consistent, the output images follow the mismatch offset accordingly.

We found that, in the cases of both mismatch in scale and spatial alignment. 4f-CNN and 4f-ONN produce consistent results. Therefore, we believe that in migrating 4f-CNN to 4f-ONN, the results of 4f-CNN can be used to actively monitor the errors introduced in building 4f-ONN, thereby improving the reliability and accuracy of 4f-ONN.This shows that our optical neural network 4f system can completely reproduce the computer neural network 4f system, and the convolution calculation in the computer is equivalently converted into the all-optical calculation of the optical neural network 4f system.

\subsection*{The influence of misaligned 4f system on image classification accuracy for optical computing}
\begin{figure}[h]
	\centering
	\includegraphics[width=\linewidth]{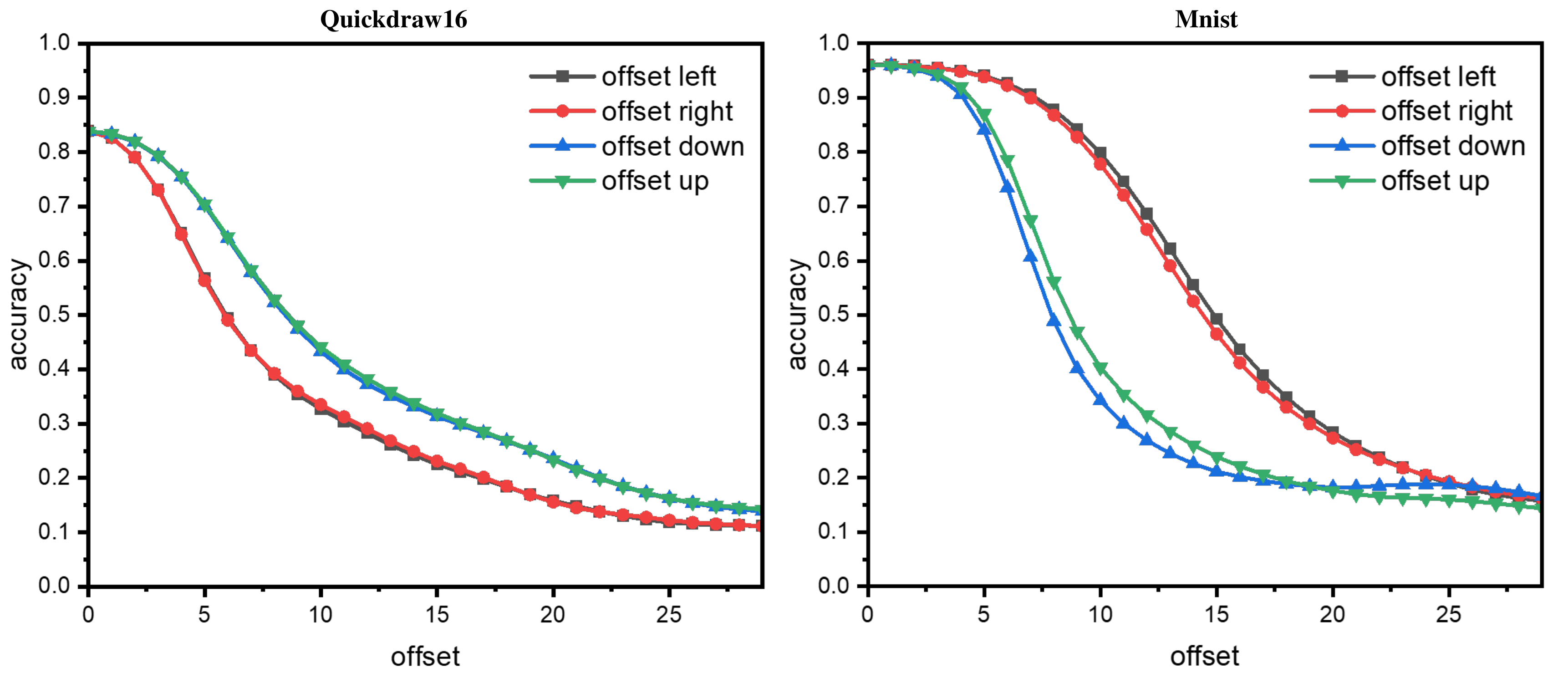}
	\caption{Classification accuracy curve with respect to spatial offset in four directions on
the Quickdraw16 and Mnist dataset}  
	\label{3}
\end{figure}

In last section, we studied the phenomenon that the output image of the 4f system can be distorted due to mismatch in scale and alignment. We further study the impact of alignment mismatch on the application of the classification task, as described in Fig. \ref{2}(a), we fixed the spatial position of the input image and shifted the SLM to the left, right, down, and up direction by 0-30 pixels,respectively. In order to comprehensively evaluate the impact of the offset, we logged every pixel translation and its corresponding prediction accuracy to draw the accuracy curve with respect to the spatial translation. Our experiment is performed on a NVIDIA GeForce GTX 1080 Ti GPU with CUDA10.1, tensorflow-gpu-2.3.0 and python-3.7.0.
In order to reduce random errors and ensure the reliability of the evaluation, we trained 100 models on the Quickdraw16 dataset for each position offset. The averaged curves of the classification accuracy with respect to offset in 4 directions are shown in Fig. \ref{3}.

As shown in Fig. \ref{3}, any offset in any direction will cause the classification accuracy of the Quickdraw16 dataset to drop.
The overall classification performance drops significantly as the offset increases, and when the offset increases to 30 pixels, the classification accuracy drops to around 10$\%$.
We found that the performance degradation was more pronounced for horizontal (left and right) offsets along the x-axis than vertical (up and down) offsets along the y-axis. We believe that the classification model is more sensitive to horizontal misalignments.

MNIST is a  database of small, square 28 × 28 pixel grayscale images of a total of 10 classes of handwritten single digits between 0 and 9, pixel values ranging from 0 to 255. The dataset contains 60,000 training images and 10,000 testing images. We only selected nine categories of numbers from 0 to 8 for our training of 4f-CNN, with about 8000 test images left after removing the number 9.  

In the same hardware and software environment setup, we also trained 100 models on the MNIST dataset and the averaged classification accuracy curve is shown in Fig. \ref{3}.
Overall, we found classification accuracy of MNIST dataset is more sensitive to both x-axis and y-axis misalignment than Quickdraw16, as the accuracy curve declines more rapidly when offset increases. While it is significantly different from the results in the Quickdraw16 dataset, the classification accuracy of the MNIST dataset is more sensitive to vertical offsets, as the classification performance degrades more significantly with respect to offsets along the y-axis. Namely, the classification accuracy of the MNIST dataset is more sensitive to vertical misalignment.

\begin{figure}[h]
	\centering
	\includegraphics[width=\linewidth]{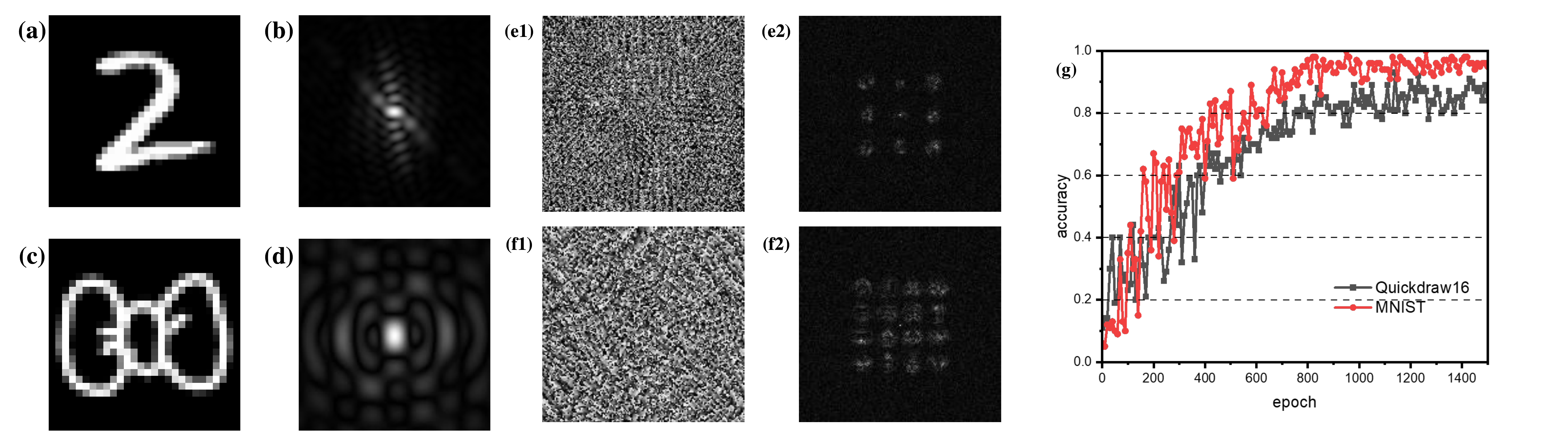}
	\caption{ (a) and (b)Digit 2 from MNIST and its spectrum at 2f.(c) and (d)Sample shape from Quickdraw16 and its spectrum at 2f.(e1)Trained phase map on MNIST.(e2)Trained PSF on MNIST.(f1)Trained phase map on Quickdraw.(f2)Trained PSF on Quickdraw.(g)Training accuracy with respect to training epochs.}
	\label{4}
\end{figure}

To explore the opposite behavior for horizontal and vertical offsets of models trained on the MNIST dataset and Quickdraw16 dataset.
We try to find the answer to the question from the semantic level of the training data. We randomly sampled two images from the MNIST dataset and the Quickdraw16 dataset, respectively, and the sampled images and their spectral light field intensity distribution at 2f are shown in Fig. \ref{4}.

The handwritten digits in MNIST are mostly spatially symmetrical on the vertical axis, and the appearance of the highlighted parts containing semantics has a significantly larger vertical distribution. Correspondingly, the vertical spectral magnitude is higher in the frequency spectrum.
Therefore, during the installation of the optical path, a small vertical offset error of the SLM may cause a substantial deformation of the optical field at 4f, resulting in a significant drop in the classification performance on the MNIST data.  
Conversely, the MNIST data has a smaller horizontal spectral field magnitude at 2f, so the distortion of the light field distribution at 4f caused by the shift in the left and right directions will be relatively weak and have a low impact on the final classification performance.

The digits in MNIST are not entirely vertical symmetry, such as the digits "4", "6", "7", we believe that the sensitivity to the vertical misalignment varies class-wisely, particularly in the testing phase.

The sketch plots in Quickdraw16 are mostly spatially symmetrical on the vertical axis. A stronger spectrum magnitude on the horizontal axis is observed in Fig. \ref{4}(c). The distortion of the light field at 4f caused by the horizontal offset error is larger, leading to a decrease in the classification accuracy.

To sum up, we believe that the difference in the sensitivity to horizontal and vertical SLM offset for the two datasets resulted from the different spectral concentrations of the mage semantics. The higher vertical density in the NMIST data leads to greater sensitivity to vertical offsets. Vice versa, Quickdraw16 data has a higher horizontal spectral density, therefore more sensitive to horizontal offsets.

To further investigate the hypothesis that the performance of 4f-CNN and 4f-ONN on different datasets has a directional dependence on image semantics. We start to observe the impact on the output image of the system at 4f when the SLM does not match the input image spectrum. 
First of all, we fine-tune two models trained on MNIST and Quickdraw. The training accuracy plots against training epochs are shown in Fig. \ref{4}(g). The MNIST dataset can reach the best training accuracy of 95$\%$ at epoch 800, and the Quickdraw16 dataset can reach 84$\%$. The corresponding phase maps and PSFs are shown in Fig. \ref{4}(e1) and Fig. \ref{4}(f1), respectively. 

In the context of the classification task, system output at 4f is followed by a mean-pooling operation to obtain the predicted class.
It can be seen that their PSFs can be visually divided into 9 blocks and 16 blocks, which are essentially the tiled 4f-CNN convolution kernels. The PSFs contain visible geometry features similar to shallow neural network convolution kernels and have the same geometric layout as the system output at 4f.

\begin{figure}[h]
	\centering
	\includegraphics[width=\linewidth]{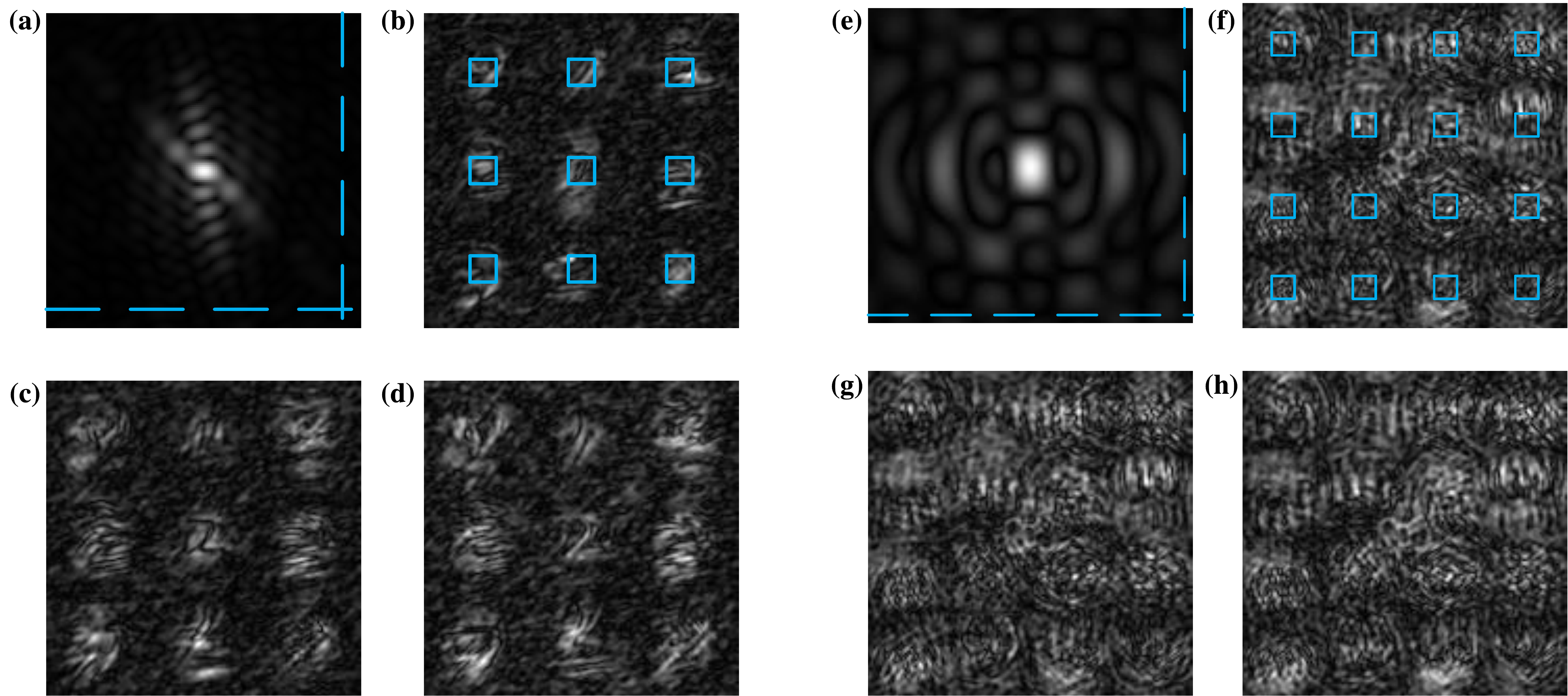}
	\caption{(a)The "2" digit image spectrum at 2f(b)output image when perfectly match SLM and input spectrum.(c)output image at 4f, 10 pixel SLM offset to the up.(d)output image at 4f, 10 pixel SLM offset to the left.(e)sample image spectrum from Quickdrow class 3.(f)system ouput when SLM perfectly matched.(g)output image at 4f, 4 pixel SLM offset to the up.(h)output image at 4f, 4 pixel SLM offset to the left.}
	\label{5}
\end{figure}

We use the images in Fig. \ref{4} as the input of the MNIST and Quickdraw16 and feed them to the trained models, respectively. 
The output images at 4f, when the SLM and the input image spectrum are perfectly matched and misaligned, are shown in Figure \ref{5}.

Fig. \ref{5}(b) is the output image at 4f of the number '2' in the MNIST dataset when the SLM perfectly matches the input image spectrum, where the blue box is the classified mean pooling region of interest(ROI). It can be clearly seen that the pixel with the highest grayscale falls within the blue box in the upper right corner, and this area corresponds to the activation of class 2, which indicates that the 4f system correctly classifies the digit '2'.
In the case of misalignment, when the SLM is shifted 10 pixels up, the output image at 4f is shown in Fig. \ref{5}(c). A significant difference is observed between this output image and the perfectly matched output in Fig. \ref{5}(b), particularly in the mean-pooling ROI. The mean-pooling activation now falls into the blue box of class 6 (lower left corner), which indicates that the 4f system classifies the digit '2' into class 6.
Similarly, we shifted SLM to the left by 10 pixels, the output image of the 4f system is shown in Fig. \ref{5}(d). The mean-pooling activation falls into class '2', although the whole image differs from the pecfectly matched SLM output. In this case, the misalignment caused distortion while the classifier still gives the correct prediction.

Regarding the Quickdraw dataset, the sample input is labeled as class 3, and the corresponding image spectrum is shown in Fig. \ref{5}(e).
Fig. \ref{5}(f) is the 4f system output image of class '3' image in the Quickdraw16 dataset with SLM matched. The trained classifier correctly predicts class 3 as the mean pooling falls into the third activation ROI in the top row. 
When the SLM is shifted up by 4 pixels,  the output image is visually very similar to the SLM matched output. The 4f system can still classify this image as class 3, as the activation falls into the same ROI.  
A significantly different output image is obtained (Fig. \ref{5}(h)) when the SLM is shifted 4 pixels to the left. In this case, the input class 3 image is misclassified as class 6 as the mean pooling activation now falls into the third ROI in the second row.

To sum up,  we verified our hypothesis that the performance of 4f-CNN and 4f-ONN have a directional dependence on image semantics. Namely, the spectral concentrations in a specific direction of the dataset will lead to classifier performance degradation when SLM is misaligned with the input image spectrum in the same direction.

\section*{Discussion}

In this work, we found that the accuracy of 4f-ONN constructed using our proposed method for light-field image processing is within $ \pm200 $ microns, which shows that our method allows SLM and the spatial position of the input image spectrum there is an error within 200 microns in a certain direction, even if there is this experimental error, the classification accuracy of the system will not be significantly reduced. The training of LFMM at 2f proposed is misalignment aware and has the potential to correct the misalignment accordingly, improving the 4f-ONN system robustness and accuracy.

\section*{Methods}

The physical process of the optical 4f system can be described by the Eq. (\ref{12121}), the output image is obtained by convolving the input image with the PSF of the optical 4f system. The conventional way to simulate optical 4f systems is to tile multiple convolution kernels superimposed in depth on a plane at a specific interval as the PSF.In terms of the classification task, the 4f system is expected to output the classification scores with multiple segments distributed spatially representing different classes. In the training process, the classification scores and ground truth labels are used to calculate the loss function and optimize the PSF of the 4f system. 
 
\begin{figure}[ht]
	\centering
	\includegraphics[width=\linewidth]{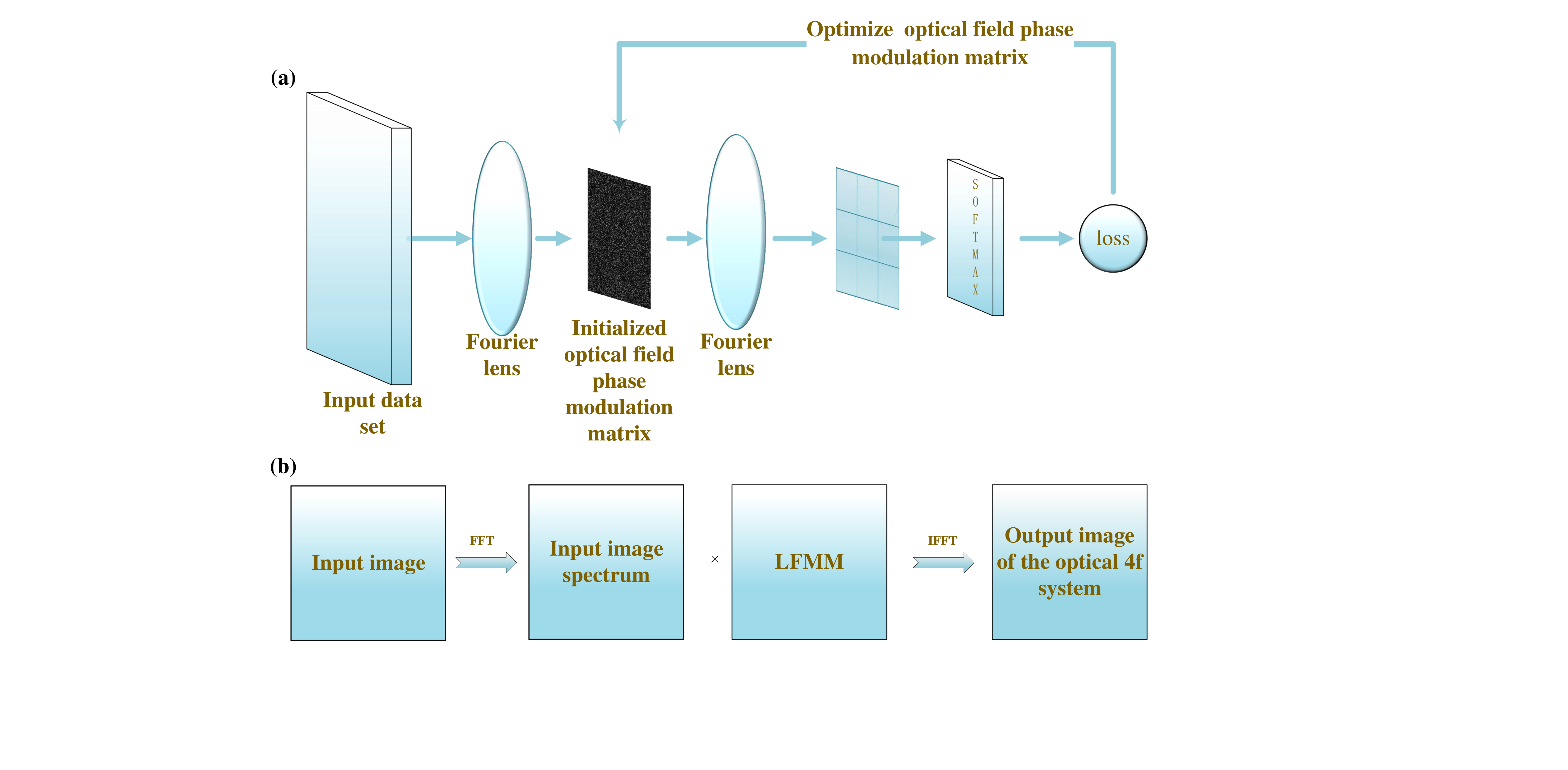}
	\caption{(a)The training of Optical neural network 4f System.(b)The 4f-ONN workflow.}
	\label{21}
\end{figure}

We first construct an LFMM for phase modulation of the light field, converting the convolution of the input image with the PSF to the multiplication of the input image spectrum with the LFMM. This process can be described by Eq. (\ref{1221}); similarly, the output image at 4f is divided into regions representing different classes as one hot classification label, and cross-entropy is performed with the standard label to obtain the loss function. Finally, back propagate the loss function to optimize the constructed LFMM to make the classification accuracy converge. The training process is shown in Fig. \ref{21}(a).

\begin{equation}
	output=input \ast  PSF
	\label{12121}
\end{equation}

\begin{equation}
	output= IFFT\left [ FFT\left ( input \right )\times LFMM \right ]
	\label{1221}
\end{equation}
The method proposed directly construct LFMM at 2f is significantly different from the use of PSF, the LFMM only interacts with the input image spectrum at 2f, eliminating the structure restriction of the PSF and achieving higher training flexibility. Consequently, the LFMM can potentially capture the image semantics feature better than the PSF in the training of classification task. 
This one-step training also gets rid of the constraints of the computer neural network. The modulation of LFMM interacts directly with the prediction output, which is very convenient to quantitatively study the LFMM and the input image spectrum mismatch caused by the inappropriate size and spatial positioning.

The system workflow is shown in Fig. \ref{21}(b). First, the Fourier transform is applied to the input image and obtains its image spectrum, and multiplied with the trained LFMM, followed by the inverse Fourier transform. Eq. (\ref{101}) calculates the spectral light field intensity of the input image at 2f:
\begin{equation}
U\left ( x,y \right )= A\left ( x,y \right )\cdot exp\left [ j\phi \left ( x,y \right ) \right ]
	\label{101}
\end{equation}

Where $ A\left ( x,y \right )$ represents the amplitude of the spectrum,$ exp\left [ j\phi \left ( x,y \right ) \right ] $is the phase of the spectrum. We do phase modulation applied only and keep amplitude constant $ A\left ( x,y \right )=1 $.We also set the amplitude of LFMM to 1 as shown in Eq. (\ref{102}):
\begin{equation}
LFMM\left ( x,y \right )= 1\cdot exp\left [ j\phi_0 \left ( x,y \right ) \right ]
	\label{102}
\end{equation}

LFMM adjusts the input image's spectrum to reconstruct its light field's spatial distribution. The output image of the optical 4f system can be obtained using the inverse Fourier transform of the reconstructed result as shown in Eq. (\ref{103}) and Eq. (\ref{104}).
\begin{equation}
output\left ( x,y \right )=IFFT\left [ U\left ( x,y \right )\cdot LFMM\left ( x,y \right ) \right ]
	\label{103}
\end{equation}
\begin{equation}
output\left ( x,y \right )=IFFT\left\{ exp\left [ j\phi \left ( x,y \right )+j\phi _0\left ( x,y \right ) \right ]\right\}
	\label{104}
\end{equation}

Similar to the popular neural network training, in a supervised learning manner. The forward pass ends after the output image of the 4f system is obtained, and the LFMM  is optimized using the back propagation algorithm. More training details will be covered in Section 3. The LFMM can also use the Eq. (\ref{105}) to calculate the PSF of an optical neural network 4f system. 

\begin{equation}
PSF\left ( x,y \right )=IFFT\left [ LFMM\left ( x,y \right ) \right ]
	\label{105}
\end{equation}

We further study the system response when the input image spectrum does not match the size and spatial position of the pre-trained LFMM. Firstly, when the input image spectrum is fixed, the 4f system response of the positional offset  $\Delta x_0$ and $\Delta y_0$ of LFMM  is calculated by the Eq. (\ref{107}).
\begin{equation}
output\left ( x,y \right )=IFFT\left [ U\left ( x,y \right )\cdot LFMM\left ( x+\Delta x_0,y+\Delta y_0 \right ) \right ]
	\label{106}
\end{equation}
\begin{equation}
output\left ( x,y \right )=IFFT\left\{ exp\left [ j\phi \left ( x,y \right )+j\phi _0\left ( x+\Delta x_0,y+\Delta y_0 \right ) \right ]\right\}
	\label{107}
\end{equation}

Similarly, the size of the input image spectrum is fixed, and the corresponding 4f system response of size mismatch with LFMM can be calculated as in Eq. (\ref{1099}).

\begin{equation}
output\left ( x,y \right )=IFFT\left [ U\left ( x,y \right )\cdot resize\left ( LFMM\left ( x,y \right ) \right ) \right ]
	\label{108}
\end{equation}
\begin{equation}
	output\left ( x,y \right )=IFFT\left\{ exp\left [ j\phi \left ( x,y \right )+resize\left ( j\phi _0\left ( x,y \right ) \right ) \right ]\right\}
	\label{1099}
\end{equation}

The physical implementation of the optical 4f system will inevitably introduce errors in optical path positioning and scale, as discussed in Eq. (\ref{107}) and Eq. (\ref{1099}). So it is necessary to quantitatively study the impact of such errors on the corresponding optical neural network prediction results. It is believed that the LFMM has better tolerance for such errors than PSF, thereby improving the classification prediction performance of the entire network system.

\bibliography{main}

\begin{thebibliography}{10}
\urlstyle{rm}
\expandafter\ifx\csname url\endcsname\relax
  \def\url#1{\texttt{#1}}\fi
\expandafter\ifx\csname urlprefix\endcsname\relax\def\urlprefix{URL }\fi
\expandafter\ifx\csname doiprefix\endcsname\relax\def\doiprefix{DOI: }\fi
\providecommand{\bibinfo}[2]{#2}
\providecommand{\eprint}[2][]{\url{#2}}

\bibitem{a1}
\bibinfo{author}{Krizhevsky, A.}, \bibinfo{author}{Sutskever, I.} \&
  \bibinfo{author}{Hinton, G.~E.}
\newblock \bibinfo{title}{Imagenet classification with deep convolutional
  neural networks}.
\newblock In \bibinfo{editor}{Pereira, F.}, \bibinfo{editor}{Burges, C.},
  \bibinfo{editor}{Bottou, L.} \& \bibinfo{editor}{Weinberger, K.} (eds.)
  \emph{\bibinfo{booktitle}{Advances in Neural Information Processing
  Systems}}, vol.~\bibinfo{volume}{25} (\bibinfo{publisher}{Curran Associates,
  Inc.}, \bibinfo{year}{2012}).

\bibitem{a2}
\bibinfo{author}{Hinton, G.} \emph{et~al.}
\newblock \bibinfo{journal}{\bibinfo{title}{Deep neural networks for acoustic
  modeling in speech recognition: The shared views of four research groups}}.
\newblock {\emph{\JournalTitle{IEEE Signal processing magazine}}}
  \textbf{\bibinfo{volume}{29}}, \bibinfo{pages}{82--97}
  (\bibinfo{year}{2012}).

\bibitem{a3}
\bibinfo{author}{LeCun, Y.}, \bibinfo{author}{Bengio, Y.} \&
  \bibinfo{author}{Hinton, G.}
\newblock \bibinfo{journal}{\bibinfo{title}{Deep learning}}.
\newblock {\emph{\JournalTitle{nature}}} \textbf{\bibinfo{volume}{521}},
  \bibinfo{pages}{436--444} (\bibinfo{year}{2015}).

\bibitem{a4}
\bibinfo{author}{Silver, D.} \emph{et~al.}
\newblock \bibinfo{journal}{\bibinfo{title}{Mastering the game of go with deep
  neural networks and tree search}}.
\newblock {\emph{\JournalTitle{nature}}} \textbf{\bibinfo{volume}{529}},
  \bibinfo{pages}{484--489} (\bibinfo{year}{2016}).

\bibitem{a5}
\bibinfo{author}{Singh, S.}, \bibinfo{author}{Okun, A.} \&
  \bibinfo{author}{Jackson, A.}
\newblock \bibinfo{journal}{\bibinfo{title}{Learning to play go from scratch}}.
\newblock {\emph{\JournalTitle{Nature}}} \textbf{\bibinfo{volume}{550}},
  \bibinfo{pages}{336--337} (\bibinfo{year}{2017}).

\bibitem{a6}
\bibinfo{author}{De~Fauw, J.} \emph{et~al.}
\newblock \bibinfo{journal}{\bibinfo{title}{Clinically applicable deep learning
  for diagnosis and referral in retinal disease}}.
\newblock {\emph{\JournalTitle{Nature medicine}}}
  \textbf{\bibinfo{volume}{24}}, \bibinfo{pages}{1342--1350}
  (\bibinfo{year}{2018}).

\bibitem{a7}
\bibinfo{author}{Silver, D.} \emph{et~al.}
\newblock \bibinfo{journal}{\bibinfo{title}{A general reinforcement learning
  algorithm that masters chess, shogi, and go through self-play}}.
\newblock {\emph{\JournalTitle{Science}}} \textbf{\bibinfo{volume}{362}},
  \bibinfo{pages}{1140--1144} (\bibinfo{year}{2018}).

\bibitem{a8}
\bibinfo{author}{Kates-Harbeck, J.}, \bibinfo{author}{Svyatkovskiy, A.} \&
  \bibinfo{author}{Tang, W.}
\newblock \bibinfo{journal}{\bibinfo{title}{Predicting disruptive instabilities
  in controlled fusion plasmas through deep learning}}.
\newblock {\emph{\JournalTitle{Nature}}} \textbf{\bibinfo{volume}{568}},
  \bibinfo{pages}{526--531} (\bibinfo{year}{2019}).

\bibitem{a9}
\bibinfo{author}{Wang, Y.} \emph{et~al.}
\newblock \bibinfo{journal}{\bibinfo{title}{Augmenting vascular disease
  diagnosis by vasculature-aware unsupervised learning}}.
\newblock {\emph{\JournalTitle{Nature Machine Intelligence}}}
  \textbf{\bibinfo{volume}{2}}, \bibinfo{pages}{337--346}
  (\bibinfo{year}{2020}).

\bibitem{a10}
\bibinfo{author}{Zikic, D.}, \bibinfo{author}{Ioannou, Y.},
  \bibinfo{author}{Brown, M.} \& \bibinfo{author}{Criminisi, A.}
\newblock \bibinfo{journal}{\bibinfo{title}{Segmentation of brain tumor tissues
  with convolutional neural networks}}.
\newblock {\emph{\JournalTitle{Proceedings MICCAI-BRATS}}}
  \textbf{\bibinfo{volume}{36}}, \bibinfo{pages}{36--39}
  (\bibinfo{year}{2014}).

\bibitem{a11}
\bibinfo{author}{Long, J.}, \bibinfo{author}{Shelhamer, E.} \&
  \bibinfo{author}{Darrell, T.}
\newblock \bibinfo{title}{Fully convolutional networks for semantic
  segmentation}.
\newblock In \emph{\bibinfo{booktitle}{Proceedings of the IEEE conference on
  computer vision and pattern recognition}}, \bibinfo{pages}{3431--3440}
  (\bibinfo{year}{2015}).

\bibitem{a12}
\bibinfo{author}{Chen, L.-C.}, \bibinfo{author}{Papandreou, G.},
  \bibinfo{author}{Kokkinos, I.}, \bibinfo{author}{Murphy, K.} \&
  \bibinfo{author}{Yuille, A.~L.}
\newblock \bibinfo{journal}{\bibinfo{title}{Deeplab: Semantic image
  segmentation with deep convolutional nets, atrous convolution, and fully
  connected crfs}}.
\newblock {\emph{\JournalTitle{IEEE transactions on pattern analysis and
  machine intelligence}}} \textbf{\bibinfo{volume}{40}},
  \bibinfo{pages}{834--848} (\bibinfo{year}{2017}).

\bibitem{a13}
\bibinfo{author}{Litjens, G.} \emph{et~al.}
\newblock \bibinfo{journal}{\bibinfo{title}{A survey on deep learning in
  medical image analysis}}.
\newblock {\emph{\JournalTitle{Medical image analysis}}}
  \textbf{\bibinfo{volume}{42}}, \bibinfo{pages}{60--88}
  (\bibinfo{year}{2017}).

\bibitem{a14}
\bibinfo{author}{Ghamisi, P.}, \bibinfo{author}{Chen, Y.} \&
  \bibinfo{author}{Zhu, X.~X.}
\newblock \bibinfo{journal}{\bibinfo{title}{A self-improving convolution neural
  network for the classification of hyperspectral data}}.
\newblock {\emph{\JournalTitle{IEEE Geoscience and Remote Sensing Letters}}}
  \textbf{\bibinfo{volume}{13}}, \bibinfo{pages}{1537--1541}
  (\bibinfo{year}{2016}).

\bibitem{a15}
\bibinfo{author}{Sokooti, H.} \emph{et~al.}
\newblock \bibinfo{title}{Nonrigid image registration using multi-scale 3d
  convolutional neural networks}.
\newblock In \emph{\bibinfo{booktitle}{International conference on medical
  image computing and computer-assisted intervention}},
  \bibinfo{pages}{232--239} (\bibinfo{organization}{Springer},
  \bibinfo{year}{2017}).

\bibitem{a16}
\bibinfo{author}{Wang, S.} \emph{et~al.}
\newblock \bibinfo{journal}{\bibinfo{title}{A deep learning framework for
  remote sensing image registration}}.
\newblock {\emph{\JournalTitle{ISPRS Journal of Photogrammetry and Remote
  Sensing}}} \textbf{\bibinfo{volume}{145}}, \bibinfo{pages}{148--164}
  (\bibinfo{year}{2018}).

\bibitem{a17}
\bibinfo{author}{Ye, F.}, \bibinfo{author}{Su, Y.}, \bibinfo{author}{Xiao, H.},
  \bibinfo{author}{Zhao, X.} \& \bibinfo{author}{Min, W.}
\newblock \bibinfo{journal}{\bibinfo{title}{Remote sensing image registration
  using convolutional neural network features}}.
\newblock {\emph{\JournalTitle{IEEE Geoscience and Remote Sensing Letters}}}
  \textbf{\bibinfo{volume}{15}}, \bibinfo{pages}{232--236}
  (\bibinfo{year}{2018}).

\bibitem{a18}
\bibinfo{author}{Dai, X.}, \bibinfo{author}{Wu, X.}, \bibinfo{author}{Wang, B.}
  \& \bibinfo{author}{Zhang, L.}
\newblock \bibinfo{journal}{\bibinfo{title}{Semisupervised scene classification
  for remote sensing images: A method based on convolutional neural networks
  and ensemble learning}}.
\newblock {\emph{\JournalTitle{IEEE Geoscience and Remote Sensing Letters}}}
  \textbf{\bibinfo{volume}{16}}, \bibinfo{pages}{869--873}
  (\bibinfo{year}{2019}).

\bibitem{a19}
\bibinfo{author}{Ardiyanto, I.} \& \bibinfo{author}{Adji, T.~B.}
\newblock \bibinfo{journal}{\bibinfo{title}{Deep residual coalesced
  convolutional network for efficient semantic road segmentation}}.
\newblock {\emph{\JournalTitle{IPSJ Transactions on Computer Vision and
  Applications}}} \textbf{\bibinfo{volume}{9}}, \bibinfo{pages}{1--5}
  (\bibinfo{year}{2017}).

\bibitem{a20}
\bibinfo{author}{Chen, L.-C.}, \bibinfo{author}{Papandreou, G.},
  \bibinfo{author}{Kokkinos, I.}, \bibinfo{author}{Murphy, K.} \&
  \bibinfo{author}{Yuille, A.~L.}
\newblock \bibinfo{journal}{\bibinfo{title}{Deeplab: Semantic image
  segmentation with deep convolutional nets, atrous convolution, and fully
  connected crfs}}.
\newblock {\emph{\JournalTitle{IEEE transactions on pattern analysis and
  machine intelligence}}} \textbf{\bibinfo{volume}{40}},
  \bibinfo{pages}{834--848} (\bibinfo{year}{2017}).

\bibitem{a21}
\bibinfo{author}{Li, B.}, \bibinfo{author}{Zhang, T.} \& \bibinfo{author}{Xia,
  T.}
\newblock \bibinfo{journal}{\bibinfo{title}{Vehicle detection from 3d lidar
  using fully convolutional network}}.
\newblock {\emph{\JournalTitle{arXiv preprint arXiv:1608.07916}}}
  (\bibinfo{year}{2016}).

\bibitem{a22}
\bibinfo{author}{Simonyan, K.} \& \bibinfo{author}{Zisserman, A.}
\newblock \bibinfo{journal}{\bibinfo{title}{Very deep convolutional networks
  for large-scale image recognition}}.
\newblock {\emph{\JournalTitle{arXiv preprint arXiv:1409.1556}}}
  (\bibinfo{year}{2014}).

\bibitem{a23}
\bibinfo{author}{Szegedy, C.}, \bibinfo{author}{Vanhoucke, V.},
  \bibinfo{author}{Ioffe, S.}, \bibinfo{author}{Shlens, J.} \&
  \bibinfo{author}{Wojna, Z.}
\newblock \bibinfo{title}{Rethinking the inception architecture for computer
  vision}.
\newblock In \emph{\bibinfo{booktitle}{Proceedings of the IEEE conference on
  computer vision and pattern recognition}}, \bibinfo{pages}{2818--2826}
  (\bibinfo{year}{2016}).

\bibitem{a24}
\bibinfo{author}{He, K.}, \bibinfo{author}{Zhang, X.}, \bibinfo{author}{Ren,
  S.} \& \bibinfo{author}{Sun, J.}
\newblock \bibinfo{title}{Deep residual learning for image recognition}.
\newblock In \emph{\bibinfo{booktitle}{Proceedings of the IEEE conference on
  computer vision and pattern recognition}}, \bibinfo{pages}{770--778}
  (\bibinfo{year}{2016}).

\bibitem{a25}
\bibinfo{author}{Huang, G.}, \bibinfo{author}{Liu, Z.}, \bibinfo{author}{Van
  Der~Maaten, L.} \& \bibinfo{author}{Weinberger, K.~Q.}
\newblock \bibinfo{title}{Densely connected convolutional networks}.
\newblock In \emph{\bibinfo{booktitle}{Proceedings of the IEEE conference on
  computer vision and pattern recognition}}, \bibinfo{pages}{4700--4708}
  (\bibinfo{year}{2017}).

\bibitem{a26}
\bibinfo{author}{Hu, J.}, \bibinfo{author}{Shen, L.} \& \bibinfo{author}{Sun,
  G.}
\newblock \bibinfo{title}{Squeeze-and-excitation networks}.
\newblock In \emph{\bibinfo{booktitle}{Proceedings of the IEEE conference on
  computer vision and pattern recognition}}, \bibinfo{pages}{7132--7141}
  (\bibinfo{year}{2018}).

\bibitem{a27}
\bibinfo{author}{Sengupta, B.} \& \bibinfo{author}{Stemmler, M.~B.}
\newblock \bibinfo{journal}{\bibinfo{title}{Power consumption during neuronal
  computation}}.
\newblock {\emph{\JournalTitle{Proceedings of the IEEE}}}
  \textbf{\bibinfo{volume}{102}}, \bibinfo{pages}{738--750}
  (\bibinfo{year}{2014}).

\bibitem{a28}
\bibinfo{author}{Duport, F.}, \bibinfo{author}{Schneider, B.},
  \bibinfo{author}{Smerieri, A.}, \bibinfo{author}{Haelterman, M.} \&
  \bibinfo{author}{Massar, S.}
\newblock \bibinfo{journal}{\bibinfo{title}{All-optical reservoir computing}}.
\newblock {\emph{\JournalTitle{Optics express}}} \textbf{\bibinfo{volume}{20}},
  \bibinfo{pages}{22783--22795} (\bibinfo{year}{2012}).

\bibitem{a29}
\bibinfo{author}{Fok, M.~P.}, \bibinfo{author}{Tian, Y.},
  \bibinfo{author}{Rosenbluth, D.} \& \bibinfo{author}{Prucnal, P.~R.}
\newblock \bibinfo{journal}{\bibinfo{title}{Pulse lead/lag timing detection for
  adaptive feedback and control based on optical spike-timing-dependent
  plasticity}}.
\newblock {\emph{\JournalTitle{Optics letters}}} \textbf{\bibinfo{volume}{38}},
  \bibinfo{pages}{419--421} (\bibinfo{year}{2013}).

\bibitem{a30}
\bibinfo{author}{Miller, D.~A.}
\newblock \bibinfo{journal}{\bibinfo{title}{Self-configuring universal linear
  optical component}}.
\newblock {\emph{\JournalTitle{Photonics Research}}}
  \textbf{\bibinfo{volume}{1}}, \bibinfo{pages}{1--15} (\bibinfo{year}{2013}).

\bibitem{a31}
\bibinfo{author}{Vandoorne, K.} \emph{et~al.}
\newblock \bibinfo{journal}{\bibinfo{title}{Experimental demonstration of
  reservoir computing on a silicon photonics chip}}.
\newblock {\emph{\JournalTitle{Nature communications}}}
  \textbf{\bibinfo{volume}{5}}, \bibinfo{pages}{1--6} (\bibinfo{year}{2014}).

\bibitem{a32}
\bibinfo{author}{Zhang, H.} \emph{et~al.}
\newblock \bibinfo{journal}{\bibinfo{title}{Integrated photonic reservoir
  computing based on hierarchical time-multiplexing structure}}.
\newblock {\emph{\JournalTitle{Optics express}}} \textbf{\bibinfo{volume}{22}},
  \bibinfo{pages}{31356--31370} (\bibinfo{year}{2014}).

\bibitem{a33}
\bibinfo{author}{Bao, X.} \emph{et~al.}
\newblock \bibinfo{journal}{\bibinfo{title}{Research progress on photoelectric
  reserve pool computing system}}.
\newblock {\emph{\JournalTitle{Advances in Lasers and Optoelectronics}}}
  \textbf{\bibinfo{volume}{52}}, \bibinfo{pages}{030005}
  (\bibinfo{year}{2015}).

\bibitem{a34}
\bibinfo{author}{Nahmias, M.~A.}, \bibinfo{author}{Tait, A.~N.},
  \bibinfo{author}{Shastri, B.~J.}, \bibinfo{author}{De~Lima, T.~F.} \&
  \bibinfo{author}{Prucnal, P.~R.}
\newblock \bibinfo{journal}{\bibinfo{title}{Excitable laser processing network
  node in hybrid silicon: analysis and simulation}}.
\newblock {\emph{\JournalTitle{Optics express}}} \textbf{\bibinfo{volume}{23}},
  \bibinfo{pages}{26800--26813} (\bibinfo{year}{2015}).

\bibitem{a35}
\bibinfo{author}{Tait, A.~N.}, \bibinfo{author}{De~Lima, T.~F.},
  \bibinfo{author}{Nahmias, M.~A.}, \bibinfo{author}{Shastri, B.~J.} \&
  \bibinfo{author}{Prucnal, P.~R.}
\newblock \bibinfo{journal}{\bibinfo{title}{Multi-channel control for microring
  weight banks}}.
\newblock {\emph{\JournalTitle{Optics Express}}} \textbf{\bibinfo{volume}{24}},
  \bibinfo{pages}{8895--8906} (\bibinfo{year}{2016}).

\bibitem{a36}
\bibinfo{author}{Shastri, B.~J.} \emph{et~al.}
\newblock \bibinfo{journal}{\bibinfo{title}{Spike processing with a graphene
  excitable laser}}.
\newblock {\emph{\JournalTitle{Scientific reports}}}
  \textbf{\bibinfo{volume}{6}}, \bibinfo{pages}{1--12} (\bibinfo{year}{2016}).

\bibitem{a37}
\bibinfo{author}{Miller, D.~A.}
\newblock \bibinfo{journal}{\bibinfo{title}{Perfect optics with imperfect
  components}}.
\newblock {\emph{\JournalTitle{Optica}}} \textbf{\bibinfo{volume}{2}},
  \bibinfo{pages}{747--750} (\bibinfo{year}{2015}).

\bibitem{a38}
\bibinfo{author}{Clements, W.~R.}, \bibinfo{author}{Humphreys, P.~C.},
  \bibinfo{author}{Metcalf, B.~J.}, \bibinfo{author}{Kolthammer, W.~S.} \&
  \bibinfo{author}{Walmsley, I.~A.}
\newblock \bibinfo{journal}{\bibinfo{title}{Optimal design for universal
  multiport interferometers}}.
\newblock {\emph{\JournalTitle{Optica}}} \textbf{\bibinfo{volume}{3}},
  \bibinfo{pages}{1460--1465} (\bibinfo{year}{2016}).

\bibitem{a39}
\bibinfo{author}{Ribeiro, A.}, \bibinfo{author}{Ruocco, A.},
  \bibinfo{author}{Vanacker, L.} \& \bibinfo{author}{Bogaerts, W.}
\newblock \bibinfo{journal}{\bibinfo{title}{Demonstration of a 4$\times$ 4-port
  universal linear circuit}}.
\newblock {\emph{\JournalTitle{Optica}}} \textbf{\bibinfo{volume}{3}},
  \bibinfo{pages}{1348--1357} (\bibinfo{year}{2016}).

\bibitem{a40}
\bibinfo{author}{Shen, Y.} \emph{et~al.}
\newblock \bibinfo{journal}{\bibinfo{title}{Deep learning with coherent
  nanophotonic circuits}}.
\newblock {\emph{\JournalTitle{Nature photonics}}}
  \textbf{\bibinfo{volume}{11}}, \bibinfo{pages}{441--446}
  (\bibinfo{year}{2017}).

\bibitem{a41}
\bibinfo{author}{Li, L.}, \bibinfo{author}{Fang, N.}, \bibinfo{author}{Wang,
  L.}, \bibinfo{author}{Huang, Z.} \emph{et~al.}
\newblock \bibinfo{journal}{\bibinfo{title}{Research progress on the
  implementation scheme of the reserve pool computing hardware}}.
\newblock {\emph{\JournalTitle{Advances in Lasers and Optoelectronics}}}
  \textbf{\bibinfo{volume}{54}}, \bibinfo{pages}{080005}
  (\bibinfo{year}{2017}).

\bibitem{a42}
\bibinfo{author}{Hughes, T.~W.}, \bibinfo{author}{Minkov, M.},
  \bibinfo{author}{Shi, Y.} \& \bibinfo{author}{Fan, S.}
\newblock \bibinfo{journal}{\bibinfo{title}{Training of photonic neural
  networks through in situ backpropagation and gradient measurement}}.
\newblock {\emph{\JournalTitle{Optica}}} \textbf{\bibinfo{volume}{5}},
  \bibinfo{pages}{864--871} (\bibinfo{year}{2018}).

\bibitem{a43}
\bibinfo{author}{Bueno, J.} \emph{et~al.}
\newblock \bibinfo{journal}{\bibinfo{title}{Reinforcement learning in a
  large-scale photonic recurrent neural network}}.
\newblock {\emph{\JournalTitle{Optica}}} \textbf{\bibinfo{volume}{5}},
  \bibinfo{pages}{756--760} (\bibinfo{year}{2018}).

\bibitem{a44}
\bibinfo{author}{Fang, M. Y.-S.}, \bibinfo{author}{Manipatruni, S.},
  \bibinfo{author}{Wierzynski, C.}, \bibinfo{author}{Khosrowshahi, A.} \&
  \bibinfo{author}{DeWeese, M.~R.}
\newblock \bibinfo{journal}{\bibinfo{title}{Design of optical neural networks
  with component imprecisions}}.
\newblock {\emph{\JournalTitle{Optics Express}}} \textbf{\bibinfo{volume}{27}},
  \bibinfo{pages}{14009--14029} (\bibinfo{year}{2019}).

\bibitem{a45}
\bibinfo{author}{Hamerly, R.}, \bibinfo{author}{Bernstein, L.},
  \bibinfo{author}{Sludds, A.}, \bibinfo{author}{Solja{\v{c}}i{\'c}, M.} \&
  \bibinfo{author}{Englund, D.}
\newblock \bibinfo{journal}{\bibinfo{title}{Large-scale optical neural networks
  based on photoelectric multiplication}}.
\newblock {\emph{\JournalTitle{Physical Review X}}}
  \textbf{\bibinfo{volume}{9}}, \bibinfo{pages}{021032} (\bibinfo{year}{2019}).

\bibitem{a46}
\bibinfo{author}{Bangari, V.} \emph{et~al.}
\newblock \bibinfo{journal}{\bibinfo{title}{Digital electronics and analog
  photonics for convolutional neural networks (deap-cnns)}}.
\newblock {\emph{\JournalTitle{IEEE Journal of Selected Topics in Quantum
  Electronics}}} \textbf{\bibinfo{volume}{26}}, \bibinfo{pages}{1--13}
  (\bibinfo{year}{2019}).

\bibitem{a47}
\bibinfo{author}{Xu, X.} \emph{et~al.}
\newblock \bibinfo{journal}{\bibinfo{title}{11 tops photonic convolutional
  accelerator for optical neural networks}}.
\newblock {\emph{\JournalTitle{Nature}}} \textbf{\bibinfo{volume}{589}},
  \bibinfo{pages}{44--51} (\bibinfo{year}{2021}).

\bibitem{a48}
\bibinfo{author}{Zhou, T.} \emph{et~al.}
\newblock \bibinfo{journal}{\bibinfo{title}{Large-scale neuromorphic
  optoelectronic computing with a reconfigurable diffractive processing unit}}.
\newblock {\emph{\JournalTitle{Nature Photonics}}}
  \textbf{\bibinfo{volume}{15}}, \bibinfo{pages}{367--373}
  (\bibinfo{year}{2021}).

\bibitem{a49}
\bibinfo{author}{Ashtiani, F.}, \bibinfo{author}{Geers, A.~J.} \&
  \bibinfo{author}{Aflatouni, F.}
\newblock \bibinfo{journal}{\bibinfo{title}{An on-chip photonic deep neural
  network for image classification}}.
\newblock {\emph{\JournalTitle{Nature}}} \bibinfo{pages}{1--6}
  (\bibinfo{year}{2022}).

\bibitem{a50}
\bibinfo{author}{Lin, X.} \emph{et~al.}
\newblock \bibinfo{journal}{\bibinfo{title}{All-optical machine learning using
  diffractive deep neural networks}}.
\newblock {\emph{\JournalTitle{Science}}} \textbf{\bibinfo{volume}{361}},
  \bibinfo{pages}{1004--1008} (\bibinfo{year}{2018}).

\bibitem{a51}
\bibinfo{author}{Luo, Y.} \emph{et~al.}
\newblock \bibinfo{journal}{\bibinfo{title}{Design of task-specific optical
  systems using broadband diffractive neural networks}}.
\newblock {\emph{\JournalTitle{Light: Science \& Applications}}}
  \textbf{\bibinfo{volume}{8}}, \bibinfo{pages}{1--14} (\bibinfo{year}{2019}).

\bibitem{a52}
\bibinfo{author}{Li, J.}, \bibinfo{author}{Mengu, D.}, \bibinfo{author}{Luo,
  Y.}, \bibinfo{author}{Rivenson, Y.} \& \bibinfo{author}{Ozcan, A.}
\newblock \bibinfo{journal}{\bibinfo{title}{Class-specific differential
  detection in diffractive optical neural networks improves inference
  accuracy}}.
\newblock {\emph{\JournalTitle{Advanced Photonics}}}
  \textbf{\bibinfo{volume}{1}}, \bibinfo{pages}{046001} (\bibinfo{year}{2019}).

\bibitem{a53}
\bibinfo{author}{Mengu, D.}, \bibinfo{author}{Luo, Y.},
  \bibinfo{author}{Rivenson, Y.} \& \bibinfo{author}{Ozcan, A.}
\newblock \bibinfo{journal}{\bibinfo{title}{Analysis of diffractive optical
  neural networks and their integration with electronic neural networks}}.
\newblock {\emph{\JournalTitle{IEEE Journal of Selected Topics in Quantum
  Electronics}}} \textbf{\bibinfo{volume}{26}}, \bibinfo{pages}{1--14}
  (\bibinfo{year}{2019}).

\bibitem{a54}
\bibinfo{author}{Qian, C.} \emph{et~al.}
\newblock \bibinfo{journal}{\bibinfo{title}{Performing optical logic operations
  by a diffractive neural network}}.
\newblock {\emph{\JournalTitle{Light: Science \& Applications}}}
  \textbf{\bibinfo{volume}{9}}, \bibinfo{pages}{1--7} (\bibinfo{year}{2020}).

\bibitem{a55}
\bibinfo{author}{Kulce, O.}, \bibinfo{author}{Mengu, D.},
  \bibinfo{author}{Rivenson, Y.} \& \bibinfo{author}{Ozcan, A.}
\newblock \bibinfo{journal}{\bibinfo{title}{All-optical information-processing
  capacity of diffractive surfaces}}.
\newblock {\emph{\JournalTitle{Light: Science \& Applications}}}
  \textbf{\bibinfo{volume}{10}}, \bibinfo{pages}{1--17} (\bibinfo{year}{2021}).

\bibitem{a56}
\bibinfo{author}{Sakib, R. M.~S.}, \bibinfo{author}{Jingxi, L.},
  \bibinfo{author}{Deniz, M.}, \bibinfo{author}{Yair, R.} \&
  \bibinfo{author}{Aydogan, O.}
\newblock \bibinfo{journal}{\bibinfo{title}{Ensemble learning of diffractive
  optical networks}}.
\newblock {\emph{\JournalTitle{Light: Science and Applications}}}
  \textbf{\bibinfo{volume}{10}} (\bibinfo{year}{2021}).

\bibitem{a57}
\bibinfo{author}{L{\'e}onard, F.}, \bibinfo{author}{Fuller, E.~J.},
  \bibinfo{author}{Teeter, C.~M.} \& \bibinfo{author}{Vineyard, C.~M.}
\newblock \bibinfo{journal}{\bibinfo{title}{High accuracy single-layer
  free-space diffractive neuromorphic classifiers for spatially incoherent
  light}}.
\newblock {\emph{\JournalTitle{Optics Express}}} \textbf{\bibinfo{volume}{30}},
  \bibinfo{pages}{12510--12520} (\bibinfo{year}{2022}).

\bibitem{a58}
\bibinfo{author}{Chang, J.}, \bibinfo{author}{Sitzmann, V.},
  \bibinfo{author}{Dun, X.}, \bibinfo{author}{Heidrich, W.} \&
  \bibinfo{author}{Wetzstein, G.}
\newblock \bibinfo{journal}{\bibinfo{title}{Hybrid optical-electronic
  convolutional neural networks with optimized diffractive optics for image
  classification}}.
\newblock {\emph{\JournalTitle{Scientific reports}}}
  \textbf{\bibinfo{volume}{8}}, \bibinfo{pages}{1--10} (\bibinfo{year}{2018}).

\bibitem{a59}
\bibinfo{author}{Yan, T.} \emph{et~al.}
\newblock \bibinfo{journal}{\bibinfo{title}{Fourier-space diffractive deep
  neural network}}.
\newblock {\emph{\JournalTitle{Physical review letters}}}
  \textbf{\bibinfo{volume}{123}}, \bibinfo{pages}{023901}
  (\bibinfo{year}{2019}).

\bibitem{a60}
\bibinfo{author}{Miscuglio, M.} \emph{et~al.}
\newblock \bibinfo{journal}{\bibinfo{title}{Massively parallel amplitude-only
  fourier neural network}}.
\newblock {\emph{\JournalTitle{Optica}}} \textbf{\bibinfo{volume}{7}},
  \bibinfo{pages}{1812--1819} (\bibinfo{year}{2020}).

\bibitem{a61}
\bibinfo{author}{Pad, P.} \emph{et~al.}
\newblock \bibinfo{title}{Efficient neural vision systems based on
  convolutional image acquisition}.
\newblock In \emph{\bibinfo{booktitle}{Proceedings of the IEEE/CVF Conference
  on Computer Vision and Pattern Recognition}}, \bibinfo{pages}{12285--12294}
  (\bibinfo{year}{2020}).

\end{thebibliography}

\section*{Author contributions statement}
Yanbing Liu and Wei Li conceived the idea.Yanbing Liu refined the idea and implemented the algorithm.Xun Liu and Wei Yang completed physics experiments under the supervision of Wei Li.Yanbing Liu wrote the manuscript under the guidance of Kun Cheng.

\section*{Additional information}
\textbf{Competing interests}: The authors declare no competing interests.

\end{document}